\documentclass[conference]{IEEEtran}
\IEEEoverridecommandlockouts
\usepackage{cite}
\usepackage{amsmath,amssymb,amsfonts}
\usepackage{algorithmic}
\usepackage{graphicx}
\usepackage{textcomp}
\usepackage{xcolor}
\def\BibTeX{{\rm B\kern-.05em{\sc i\kern-.025em b}\kern-.08em
    T\kern-.1667em\lower.7ex\hbox{E}\kern-.125emX}}

\newcommand{\RO}{\texttt{ibmq\_rochester}} 

\graphicspath{{figures/}}

\begin{document}
\title{Characterizing the memory capacity of\\ transmon qubit reservoirs\\
\thanks{This manuscript has been authored by UT-Battelle, LLC under Contract No. DE-AC05-00OR22725 with the U.S. Department of Energy. The United States Government retains and the publisher, by accepting the article for publication, acknowledges that the United States Government retains a non-exclusive, paid-up, irrevocable, world-wide license to publish or reproduce the published form of this manuscript, or allow others to do so, for United States Government purposes. The Department of Energy will provide public access to these results of federally sponsored research in accordance with the DOE Public Access Plan
(http://energy.gov/downloads/doe-public-access-plan).}
}

\author{Samudra~Dasgupta$^{1,2}$, Kathleen~E.~Hamilton$^2$, Arnab~Banerjee$^3$\\\\
\textit{$^1$Bredesen Center, University of Tennessee, USA}\\
\textit{$^2$Quantum Computational Science, Oak Ridge National Laboratory, USA}\\
\textit{$^3$Department of Physics and Astronomy, Purdue University, USA}\\
}

\maketitle

\begin{abstract}
Quantum Reservoir Computing (QRC) exploits the dynamics of quantum ensemble systems for machine learning. Numerical experiments show that quantum systems consisting of 5–7 qubits possess computational capabilities comparable to conventional recurrent neural networks of 100 to 500 nodes. Unlike traditional neural networks, we do not understand the guiding principles of reservoir design for high-performance information processing. Understanding the memory capacity of quantum reservoirs continues to be an open question. In this study, we focus on the task of characterizing the memory capacity of quantum reservoirs built using transmon devices provided by IBM. Our hybrid reservoir achieved a Normalized Mean Square Error (NMSE) of $6 \times 10^{-4}$ which is comparable to recent benchmarks. The Memory Capacity characterization of a $n$-qubit reservoir showed a systematic variation with the complexity of the topology and exhibited a peak for the configuration with $n-1$ self-loops. Such a peak provides a basis for selecting the optimal design for forecasting tasks.
\end{abstract}

\begin{IEEEkeywords}
Quantum Reservoir Computing, Memory Capacity, Time-series forecasting, Data Science
\end{IEEEkeywords}

\section{Introduction}\label{sec:intro}
The field of classical reservoir computing (RC) \cite{gerstner2014neuronal} provides a road map towards using signal-driven dynamical systems to process information with non-von Neumann architectures. RC models are useful in providing alternatives to deep learning that can deliver comparable performance yet are low energy, and computationally simple. They are capable of both one-shot and continuous real-time learning and excel at non-linear function approximation tasks. Quantum Reservoir Computing (QRC) is a natural extension that exploits the quantum dynamics of ensemble systems for information processing (specifically machine learning). The key is to find an appropriate form of physics that exhibits rich dynamics, thereby allowing us to outsource a part of the computation.
\par
RC systems have been utilized in many different applications and can be constructed from many different dynamical systems (see recent reviews in \cite{dambre2012information, tanaka2019recent}). There have been several approaches to quantum reservoir designs and numerical experiments show that quantum systems consisting of 5--7 qubits possess computational capabilities comparable to conventional recurrent neural networks of 100 to 500 nodes \cite{fujii2017harnessing}. Additionally, small quantum systems also demonstrate significant computational capacities \cite{govia2020quantum}. There have been several applications of QRC most notably time-dependent signal processing, speech recognition, NLP, sequential motor control of robots, and stock market predictions. A recent study \cite{kutvonen2020optimizing} has focused on optimizing quantum reservoirs for time series forecasting for financial data (the S\&P 500 index). In another recent work \cite{nakajima2019boosting}, quantum spin systems were used to construct a quantum reservoir and used for predicting non-linear time series. Reservoirs built using superconducting qubits are demonstrated in \cite{chen1,chen2}.
\par
Our motivation is to implement systematic design considerations of hybrid quantum-classical reservoirs that incorporate Noisy Intermediate-Scale Quantum (NISQ) hardware and to present results which should be useful for practitioners. Memory capacity (MC) quantifies the ability of a reservoir to forecast at different time-scales (see Section \ref{sec:ps}). We denote our hybrid reservoirs "NISQ reservoirs" and present results on their MC characterization for NISQ reservoirs which incorporate quantum circuits and classical feedback elements (detailed in Section \ref{sec:res_design}). We address the question of evaluating the MC of various reservoir topologies and how to select the optimal one. We characterize the MC of different NISQ reservoir configurations executed on IBM quantum hardware (Section \ref{sec:NISQ_capacity}).  Our methods follow the approach given in \cite{nakajima2019boosting}. Our approach is also comparable to \cite{chen1, chen2}. The NISQ reservoir configuration with the highest MC is then benchmarked using a time-series prediction task (Section \ref{sec:benchmarking}). 

\section{Memory characterization}\label{sec:ps}
Let $u_k$ be the time-series one is trying to forecast (where $k$ denotes the time index).
Let $\hat{u}_{k-\tau}$ denote the forecast of $\hat{u}_{k}$ using information till time-step $k-\tau$. The correlation $r_\tau$ between $\hat{u}_{k-\tau}$ and $u_k$ is a measure of how well the system is able to do a $\tau$ step look-ahead prediction: 
\begin{equation}
r_\tau^2 = \frac{ \mathrm{COV}(u_{k-\tau}, \hat{u}_{k-\tau})}{\sigma^2(u_{k-\tau}) \sigma^2(\hat{u}_{k-\tau})},
\end{equation}
where COV(x,y) denotes the covariance between x and y and $\sigma(x)$ denotes the standard deviation of x. Intuitively, one expects that the larger the value of $\tau$, the lower is the value of $r_\tau$ (as higher the value of $\tau$, more amount of recent data is ignored).

The MC is the sum of $r_\tau^2$ over different values of $\tau$:
\begin{equation}
MC = \sum\limits_{\tau=1}^{\tau_{max}} r_\tau^2.
\label{eq:mc_def}
\end{equation}
As in \cite{nakajima2019boosting}, we use a random sequence of values in the range $\in [0, 1]$ for $u_k$ (where $k$ denotes the time index) and fix the maximum value of $\tau$ to be $\tau_{max}=120$. This is done to ensure that the MC benchmark does not depend on a specific time-lag or a specific signal structure.

\section{NISQ Reservoir Design}
\label{sec:res_design}
\begin{figure}[!htbp]
\centering
\includegraphics[width=0.9\columnwidth]{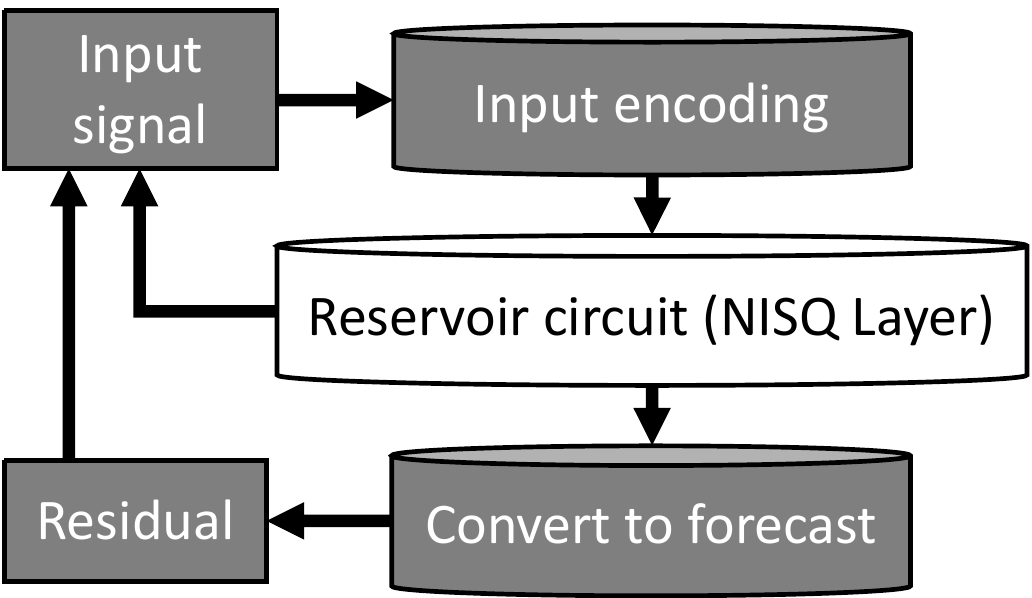}
\caption{Schematic of the hybrid quantum-classical reservoir (NISQ reservoir) system which consists of classical inputs and outputs (grey boxes),  classical computational layers (grey cylinders) and quantum computational layers (white cylinder).}
\label{fig:schematic}
\end{figure}

The NISQ reservoirs used in this study are hybrid quantum-classical systems.  The demarcation between classical and quantum resources is shown in Fig.~\ref{fig:schematic}. The first classical layer transforms the input into a qubit angle encoding. The quantum layer is used to generate an array of N-qubit spin values. The final classical layer is used to compute the forecast, and the forecast error. Both the forecast error and spin values are fed back into the first classical layer.
\par
We characterize the MC of a N-qubit NISQ reservoir as a function of recurrent connections using a sequence of $1+N+\frac{N(N-1)}{2}$ graphs in increasing order of network connectivity (and hence complexity). The first term in the sequence is an empty graph on $N$ vertices.  The next $N$ terms in the sequence are sequentially constructed by adding self-loops to each vertex.  The next $N$ terms are sequentially constructed by connecting the $N$ vertices into a simple cycle.  Finally the remaining ($\frac{N(N-1)}{2}$) terms of the sequence are constructed by sequentially connecting vertices until the final circuit is a fully connected graph with $N$ self-loops. Note that an edge can be realized between any two nodes of the reservoir if a two-qubit gate is placed between the qubits in the quantum layer; or if the output of one qubit is fed to another qubit during the classical pre-processing layer.
\begin{figure}[htbp]
\centering
\includegraphics[width=3.3in]{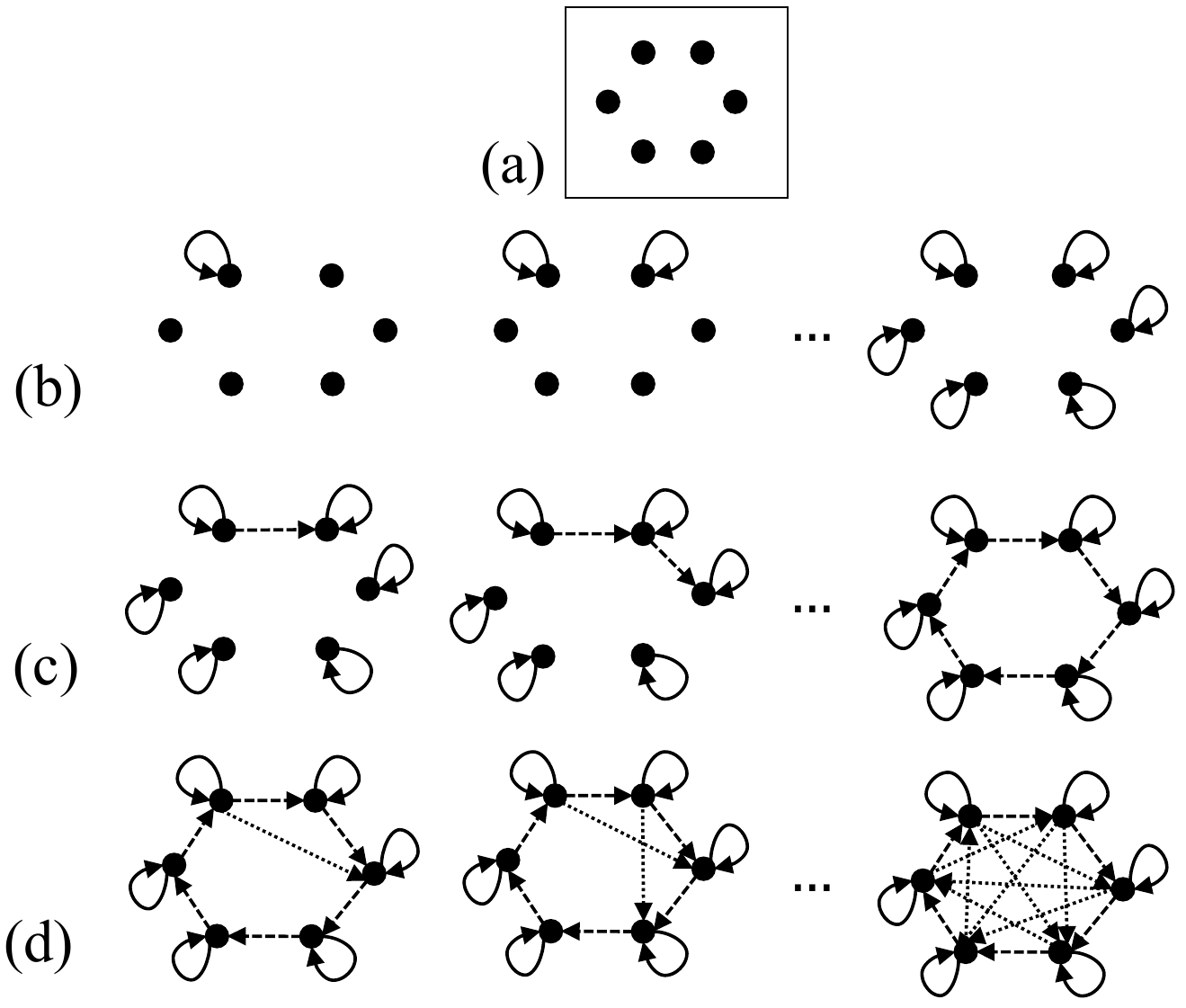}
\caption{Sequence of reservoir complexity circuits: (a) The first term is always an empty graph on $N$ qubits, (b) The first ($N$) circuits are generated by adding self-loops, (c) The next ($N$) circuits are generated by connecting the qubits into a simple cycle, (d) The remaining circuits are generated by adding edges to fully connect all $N$ qubits.}
\label{fig:bath_graph_sequence}
\end{figure}

\section{MC Characterization on NISQ Hardware}
\label{sec:NISQ_capacity}
\begin{figure}[htbp]
\centering
\includegraphics[width=3.3in]{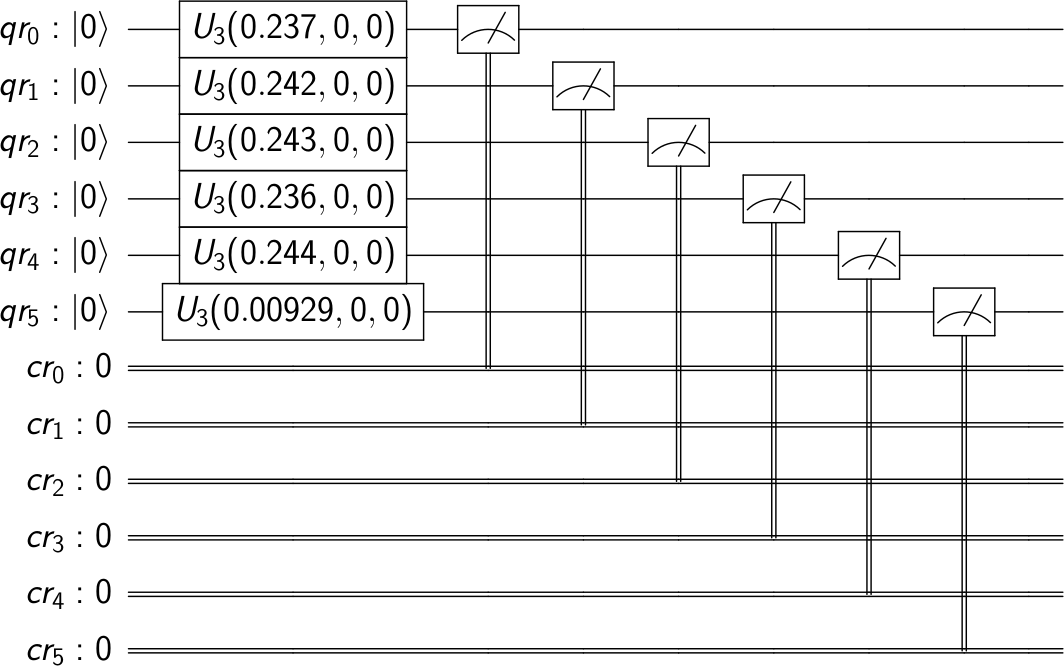}
\caption{One of the 22 configurations executed on \RO{} (a $53$ qubit platform that has been recently retired by IBM) using six $U3(\theta,\phi,\lambda)$ gates. The qubits were selected based on the lowest error rates at the time of job execution. Each circuit was sampled using $8192$ shots.}
\label{fig:qrc_circuit.png}
\end{figure}

\begin{figure}
\centering
\includegraphics[width=3.3in]{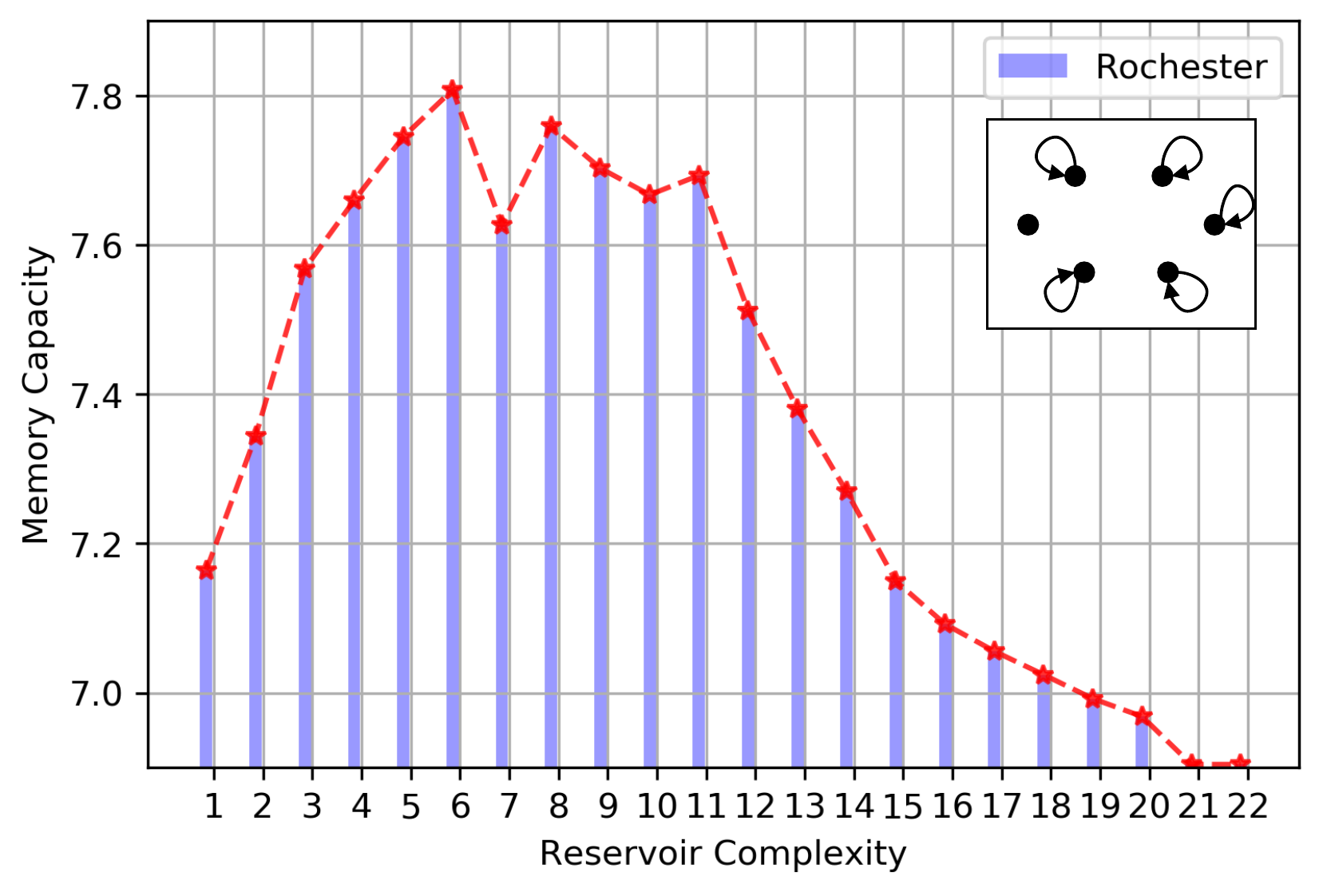}
\caption{MC evaluated with a random input signal as a function of reservoir complexity for a 6-qubit reservoir executed on \RO. [Inset] The optimal reservoir topology with self-loops on 5 qubits.}
\label{fig:mc_6qubits_apr4_inset}
\end{figure}

\begin{figure}
\centering
\includegraphics[width=\columnwidth]{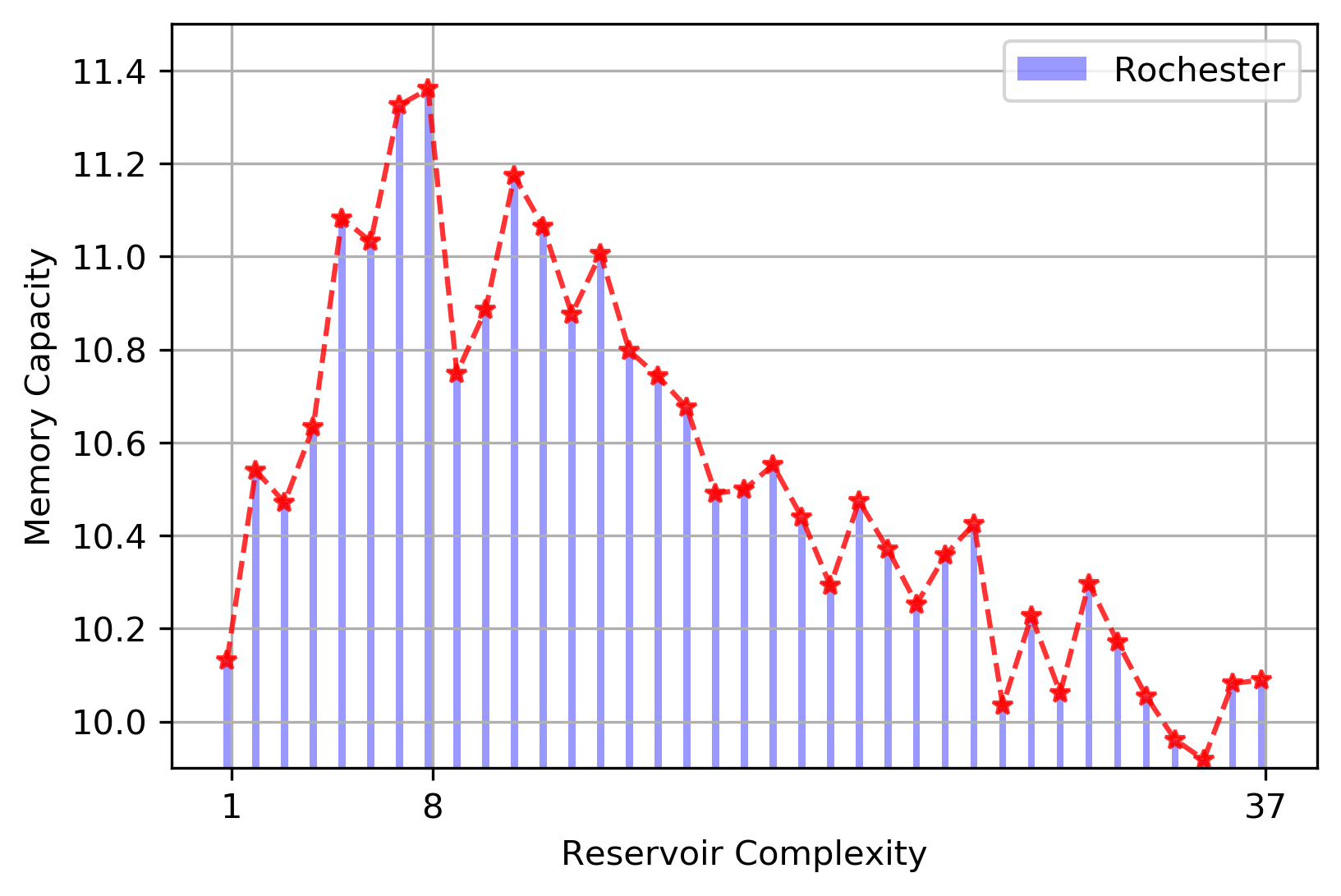}
\caption{MC evaluated with a random input signal as a function of reservoir complexity for a 8-qubit reservoir executed on \RO.}
\label{fig:mc_8qubits_apr12}
\end{figure}

\begin{figure}[htbp]
\centering
\includegraphics[width=3.3in]{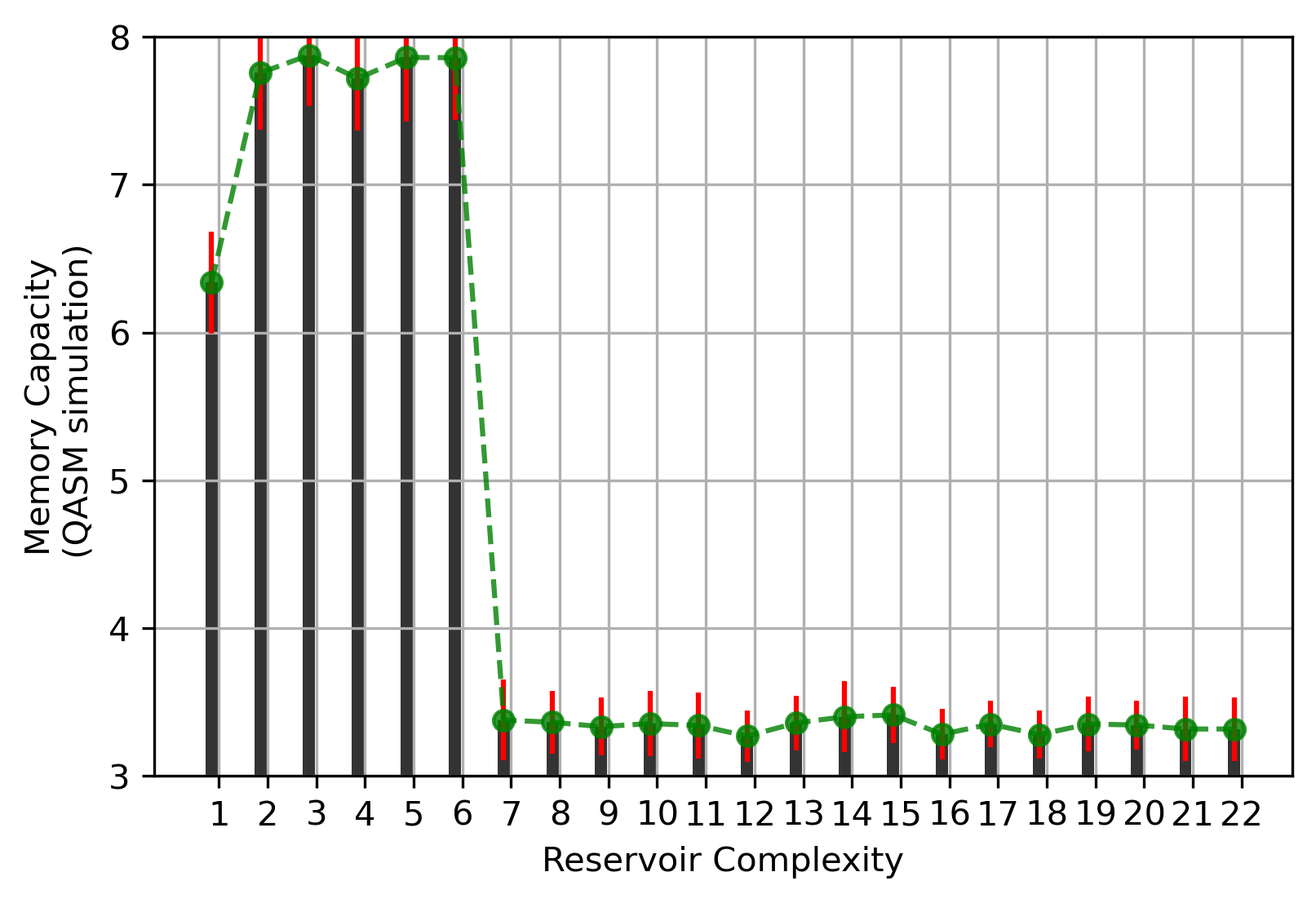}
\caption{MC as a function of reservoir complexity for a 6-qubit reservoir simulated with noiseless qubits.}
\label{fig:mc_6qubits_mar07_2020}
\end{figure}

MC characterization of a 6 qubit hybrid reservoir (shown in Fig.~\ref{fig:mc_6qubits_apr4_inset}) was evaluated using IBM's $52$ qubit platform (\RO). The 22 configurations possible are shown in Fig.~\ref{fig:bath_graph_sequence}. The quantum circuit for one of the 22 configurations is shown in Fig. \ref{fig:qrc_circuit.png}
The optimal reservoir topology has self-loops on 5 qubits (i.e. a peak in MC, within the bounds of statistical significance, is observed for reservoirs with 5 self-loops). MC characterization for an 8 qubit hybrid reservoir is shown in Fig.~\ref{fig:mc_8qubits_apr12}. There are $36$ possible graphs and the peak memory capacity is observed for the reservoir with $n-1 = 7$ self-loops.
The same sequence of reservoir topologies were also simulated in IBM Qiskit \cite{Qiskit}. The results of the noiseless simulation are shown in Fig.~\ref{fig:mc_6qubits_mar07_2020}. Comparison with Fig.~\ref{fig:mc_6qubits_apr4_inset} reveals that the hardware noisiness translates into higher MC (within the bounds of statistical significance) for circuits with higher connectivity (leading to higher degree of non-linear dynamics). We also observe a slower decay in MC for the NISQ reservoir with hardware noise. This points to a beneficial impact of the noise in today's NISQ devices.
\section{Application Benchmarking}
\label{sec:benchmarking}
The Non-linear Auto-regressive Moving Average (NARMA) series is a forecasting task that is commonly employed as a performance benchmark.  It has a high degree of non-linearity and dependence on long time lags, leading to significant memory requirements in the forecasting model. We use one step ahead forecasting of the NARMA5 series to benchmark the performance of our quantum reservoir construction.  This benchmark was executed using simulated noisy qubits with the noise modeling capabilities available in Qiskit \cite{Qiskit}. The NARMA5 series is a temporal sequence defined by:
\begin{equation}
\begin{split}
v_{t+1} = &\alpha v_t + \beta v_t (v_t + v_{t-1} + v_{t-2} + v_{t-3} + v_{t-4}) + \\
&\gamma s_{t-4}s_t + \delta,\\
s_t = &\mu \left[ sin\frac{2\pi f_0 t}{T} sin\frac{2\pi f_1 t}{T} sin\frac{2\pi f_2 t}{T} + 1\right].
\end{split}
\label{eq:NARMA_eqs}
\end{equation}

The parameters in Eq. \ref{eq:NARMA_eqs} are: $\alpha = 0.30, \beta = 0.05, \gamma = 1.50, \delta = 0.10, \mu = 0.10$, and  $f_0 = 2.11,f_1 = 3.73, f_2 = 4.11, T = 100$.  These values were originally used in \cite{fujii2017harnessing} to benchmark quantum reservoirs.
\begin{figure}[htbp]
\centering
\includegraphics[width=\columnwidth]{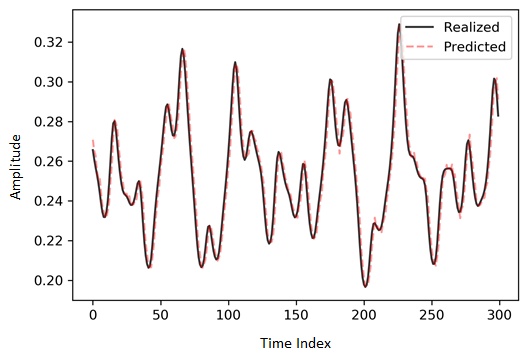}
\caption{One-step ahead predictions for the NARMA-5 time-series with the quantum reservoir executed with noisy simulation in Qiskit.}
\label{fig:narmaPred}
\end{figure}
\begin{figure}
\centering
\includegraphics[width=\columnwidth]{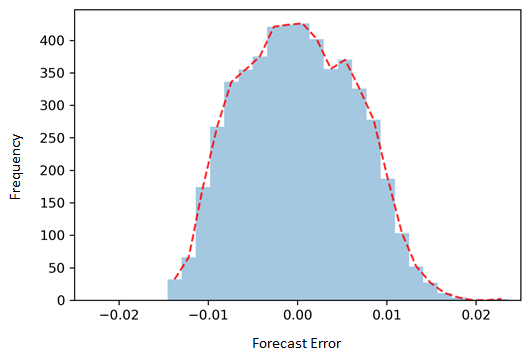}
\caption{Histogram of normalized mean square error for the NARMA5 prediction task.}
\label{fig:narmaErrHist.png}
\end{figure}

Fig.~\ref{fig:narmaPred} shows the comparison of realized vs predicted time-series for the NARMA5 task. Only a zoomed-in snapshot is shown of the 5000 point long sequence. The initial one-third of the data was flushed out to allow the system to stabilize. Our hybrid reservoir achieved an NMSE of $6\times 10^{-4}$. One can compare this to the NMSE obtained in \cite{fujii2017harnessing} which lied in the range $[3\times 10^{-3}, 7.6\times 10^{-6}]$. Thus, the benchmark performance of our hybrid reservoir is comparable to the benchmark performance found in \cite{fujii2017harnessing}. We observe low bias in the prediction error (see Fig. \ref{fig:narmaErrHist.png}).
\section{Reservoir Characteristics}
\label{sec:discussion}
In this section we discuss important reservoir characteristics for time-series forecasting. RC relies on a reservoir of randomly connected oscillators in which the connections are not trained. It uses a simple readout that is suited to low-cost, real-time, history-dependent, dynamical responses to external inputs. The inputs are mapped to a high dimensional space denoted by the reservoir state vector $\mathbf{x}(n)$, where:\\
\begin{equation}
\mathbf{x}(n) = 
\begin{bmatrix}
    x_0(n)   \\
    x_{1}(n) \\
    \vdots   \\    
    x_{N-1}(n)
\end{bmatrix}
\end{equation} 
and each $x_i$ represents the state of a node in the reservoir. The output from the high dimensional space is trained to predict the desired function using a simple method (like linear regression).
\begin{enumerate}
\item Common Signal Induced Synchronization: If the reservoir has two different initial state $s(t_0)$ and $\hat{s}(t_0)$, then, if provided with the same input stimuli $\{u(t)\}_{t\geq t_0}$,  it must satisfy,
\begin{equation}
|| s(t) - \hat{s}(t)|| \rightarrow 0 \textrm{ as } t \rightarrow \infty.   
\end{equation}
Another way of stating this is that the reservoir must have fading memory (also know as echo state property in literature): the outputs of the dynamical system should stay close if the corresponding input are close in recent times \cite{inubushi2017reservoir}. This can be viewed as a consistency or convergence criterion, it ensures that any computation performed by the reservoir is independent of its initial condition.

\item Reservoir Dimensionality: A reservoir should have adequate (preferably exponential in number of nodes) linearly independent internal variables. The number of linearly independent variables of the NISQ reservoir (the Hilbert space dimension) gives an upper limit on the computational capacity. As noted in \cite{ghosh2019quantum} prediction accuracy improves as you increase the number of nodes in the system. 

\item Adequate Memory: A reservoir can have memory of past inputs \cite{farkas2016computational}. Using a one qubit reservoir for simplicity, let's understand how memory manifests in a dynamical system. Suppose $u(t)$ and $\hat{u}(t)$ are two identical time series, except for a small perturbation at $ t = t_0-1$:
\begin{equation*}
\begin{split}
&\hat{u}(t_0-1) = u(t_0-1) + \Delta \textrm{, for } t = t_0 - 1,\\
&\hat{u}(t) = u(t) \textrm{, for all } t \neq t_0 - 1.
\end{split}
\end{equation*}
When we feed $u(t)$ or $\hat{u}(t)$ into the quantum circuit, we get the spin time series $\{s(t)\}$ and $\{\hat{s}(t)\}$ respectively. If $\delta s(t) = s(t) - \hat{s}(t)$ denotes the difference between the outputs $s(t)$ and $\hat{s}(t)$, then we say the reservoir has memory when $\delta s(t)$ and $\delta s(0)$ are related (i.e. $\delta s(t)$ can provide information about $\delta s(0)$). Higher mutual information between $\delta s(t)$ and $\delta s(0)$ implies higher MC.  A formal proof is given in \cite{inubushi2017reservoir}.  A linear circuit has higher MC as $\delta s(t)$ is strongly correlated with $\delta s(0)$. Thus high degree of linearity is more suitable for forecasting tasks which need to recall historical patterns. This implies that to introduce linear elements in the NISQ reservoir we will need to introduce `self-loops' in the spin-system.

\item Response Separability: The separation property is the reservoir's capability to generate dynamics sufficiently rich that can can distinguish between any two different input sequences. This is important because it is not enough that the reservoir is excitable by the input sequence you care about. It should be excitable by any distinguishable inputs and the (input history dependent) response should be adequately distinguishable \cite{tanaka}.

\item Adequate Non-linearity: Non-linearity is required for effective functioning of reservoir computers to address the \textit{linearly inseparable problem} \cite{kia2017nonlinear}. A non-linear transformation is mandatory for tasks such as classification by support vector machines. This property turns out to be crucial for achieving universal computing. However, non-linearity also degrades memory. Thus a careful trade-off is required between the linear and non-linear elements of the circuit. 

\item Edge Density: Edge density is a system level metric (as opposed to node level metric) that is an important driver of the predictive power achieved by a hybrid reservoir. We quantitatively define edge density as the ratio of the total number of edges present in the reservoir configuration to the total number of possible edges. A discussion on how heightened non-linearity in the system due to increased connectivity leads to MC degradation can be found in \cite{inubushi2017reservoir}. 

\item Feedback Strength: To be an effective forecasting engine, the reservoir has to strike a balance between two competing aims: memorizing past patterns (which is related to over-fit reduction) and reducing mean square error (which is related to fit accuracy). The former requirement asks for the `state signal' to play a dominant role (as the reservoir memorizes through the time evolution of its quantum spin state) while the latter pushes the `incoming signal pattern' to have more weighting. This tunable parameter can be used in the system evolution specification.

\item Noise induced regularization: It is well-known that it is possible to use dissipative quantum systems as universal function approximators for temporal information processing even in the presence of noise. Such noise can be beneficial in machine learning related information processing tasks. It plays a role akin to regularization \cite{noh2017regularizing}. The phrase `to regularize' means `to make more acceptable'. Function approximators become more acceptable when they `train' on 'noisy' data and thereby avoid over-fitting. Thus noise induced regularization helps NISQ reservoirs to be `well-behaved' and avoid taking extreme values in forecasting related tasks.

\end{enumerate}
Some progress has been made in developing a theoretical underpinning behind the ability to use dissipative quantum systems as the quantum counterpart to approximating non-linear input-output maps using classical dynamical systems. However, understanding the computational capacity of quantum reservoirs continues to be an open question.

\section{Conclusion}
Quantum Reservoir Computing (QRC) does not require sophisticated quantum gate (natural dynamics is enough) and thus, exhibits high feasibility (although the scope of application seems restricted to data-fitting tasks). Numerical experiments show that quantum systems consisting of 5–7 qubits possess computational capabilities comparable to conventional recurrent neural networks of 100 to 500 nodes \cite{fujii2017harnessing}.
\par
In this study, we characterized and benchmarked the memory capacity of various noisy, transmon qubit reservoir topologies. Our hybrid reservoir achieved a Normalized Mean Square Error (NMSE) of $6 \times 10^{-4}$ which is comparable to recent benchmarks. The Memory Capacity characterization of a $n$-qubit reservoir showed a systematic variation with the complexity of the topology and exhibited a peak for the configuration with $n-1$ self-loops. Such a peak provides a basis for selecting the optimal design for forecasting tasks.
\par
Promising avenues of future work include analyzing the QRC performance for specific $\tau$-step look-ahead predictors. We are currently working on evaluating the efficacy of QRC in predicting financial time-series data (such as VIX) and modeling the noisy quantum dynamics accurately to understand the sources of non-linearity.

\section*{Acknowledgements}
This research used quantum computing resources of the Oak Ridge Leadership Computing Facility, which is a DOE Office of Science User Facility supported under Contract DE-AC05-00OR22725. This work was partially supported as part of the ASCR QCAT program at Oak Ridge National Laboratory under FWP \#ERKJ347. This work was partially supported as part of the ASCR Fundamental Algorithmic Research for Quantum Computing Program at Oak Ridge National Laboratory under FWP ERKJ354. Part of the support for SD and AB came from College of Science, Purdue University.
\bibliographystyle{ieeetr}
\bibliography{qrc_refs.bib}

\end{document}